# Analysis of multiplicity dependencies of midrapidity $p_t$ distributions of identified charged particles in $p+p$ collisions at $(s)^{1/2}$=7 TeV at the LHC

Khusniddin K. Olimov[1*], Fu-Hu Liu[2#], Kobil A. Musaev[1], Maratbek Z. Shodmonov[1]

[1]Physical-Technical Institute of Uzbekistan Academy of Sciences, Chingiz Aytmatov str. 2[b], 100084 Tashkent, Uzbekistan
[2]Institute of Theoretical Physics & Collaborative Innovation Center of Extreme Optics & State Key Laboratory of Quantum Optics and Quantum Optics Devices, Shanxi University, Taiyuan 030006, China

[*]khkolimov@gmail.com, kh.olimov@uzsci.net
[#]fuhuliu@sxu.edu.cn

**Abstract.** Multiplicity dependencies of midrapidity $p_t$ distributions of identified charged particles in inelastic proton-proton ($p+p$) collisions at center-of-mass energy $(s)^{1/2}$=7 TeV at the Large Hadron Collider (LHC), measured by ALICE Collaboration, have been analyzed. The combined minimum $\chi^2$ fits with thermodynamically consistent Tsallis function as well as Hagedorn function with the embedded transverse flow have described quite well the $p_t$ spectra of the charged pions and kaons, protons and antiprotons in the studied ten different classes of charged-particle multiplicity density. The extracted effective temperatures $T$ of thermodynamically consistent Tsallis function have demonstrated consistent rise with increasing the multiplicity of charged particles in $p+p$ collisions at $(s)^{1/2}$=7 TeV in agreement with the similar result obtained recently in $p+p$ collisions at $(s)^{1/2}$=13 TeV at the LHC. The corresponding $T$ versus $<dN_{ch}/d\eta>$ dependence in inelastic $p+p$ collisions at $(s)^{1/2}$=7 TeV is reproduced quite well by the simple power function with the same value (≈ 1/3) of exponent parameter as that extracted in inelastic $p+p$ collisions at $(s)^{1/2}$=13 TeV. The same power dependence $T \sim \varepsilon^{1/3}$ between the energy density and effective temperature of the system is observed in inelastic $p+p$ collisions at $(s)^{1/2}$=7 and 13 TeV. It is found that the transverse (radial) flow emerges at $<dN_{ch}/d\eta>$ ≈ 6 and then increases, becoming significant at higher multiplicity events in $p+p$ collisions at $(s)^{1/2}$=7 TeV. It is estimated from analysis of $T_0$ and $\langle\beta_t\rangle$ versus $<dN_{ch}/d\eta>$ dependencies that the probable deconfinement phase transition in $p+p$ collisions at $(s)^{1/2}$=7 TeV occurs at $<dN_{ch}/d\eta>$ ≈ 6.1±0.3, which is noticeably smaller of the corresponding recent estimate ($<dN_{ch}/d\eta>$ ≈ 7.1±0.2) in $p+p$ collisions at $(s)^{1/2}$=13 TeV. The corresponding critical energy densities for probable deconfinement phase transition in $p+p$ collisions at $(s)^{1/2}$=7 and 13 TeV at the LHC have been estimated to be 0.67±0.03 GeV/fm$^3$ and 0.76±0.02 GeV/fm$^3$, respectively.

*Keywords*: transverse momentum distributions; collective phenomena in high-energy proton-proton ($p+p$) collisions at the LHC; non-extensive Tsallis distribution function; non-extensivity parameter $q$; Hagedorn function with the embedded transverse flow; transverse (radial) flow velocity; effective temperature; kinetic freeze-out temperature; deconfinement phase transition.

## 1. Introduction

The non-extensive Tsallis distribution functions have been quite useful and efficient in description of the transverse momentum ($p_t$) distributions of the final particles in high-energy collisions [1-21]. There exist different versions of Tsallis distribution function, which have described equally well the $p_t$ spectra of final hadrons in $p+p$ collisions, including the largest measured $p_t$ values at the Relativistic Heavy Ion Collider (RHIC) and Large Hadron Collider (LHC) experiments [3-10, 15]. The Tsallis functions have parameterized quite well the measured $p_t$ distributions of particles up to 200 GeV/$c$ in $p+p$ collisions at center-of-mass energy $(s)^{1/2}$=7 TeV at the LHC [9]. The non-extensivity parameter $q$ of Tsallis function has shown quite noticeable sensitivity to the high transverse momentum region ($p_t$>3 GeV/$c$) of the invariant $p_t$ spectra of particles, implying importance and necessity of analyzing the long $p_t$ ranges for extraction of the more precise $q$ values [22-24].

The wide use of Tsallis distribution function is due to its excellent parameterization of the experimental $p_t$ distributions of particles using just three parameters: one parameter representing the effective temperature ($T$) of a system, the second one, non-extensivity parameter $q$, which accounts for the degree of deviation of $p_t$ distribution from the Boltzmann-Gibbs exponential distribution, and the third one - the fitting (normalization) constant, thought to be proportional to the system volume. It is important to note that the Tsallis distribution has the advantage of being connected (via the entropy) to thermodynamics, which is not the case for other power law distributions [8]. Besides providing information on deviation from the extensive Boltzmann-Gibbs statistics, the fit parameters $q$ and $T$ can also be used for identification of the initial conditions and system size scaling [25].

To describe the $p_t$ distributions of particles produced in high-energy heavy-ion as well as $p+p$ collisions at the RHIC and LHC, various transverse flow models have been incorporated into Tsallis statistics. To estimate the kinetic freeze-out temperature and transverse expansion velocity, mostly the Blast-Wave model with Boltzmann Gibbs statistics (the BGBW model) [26-28], the Blast-Wave model coupled with Tsallis statistics (the TBW model) [29,30], the Tsallis distribution with transverse flow effect – Improved Tsallis distribution [30-32], and Hagedorn formula (function) with the embedded transverse flow [11,16,21,33] have been used.

In most of previous analyses of high energy $p+p$ and heavy-ion collisions, the various versions of Tsallis function, also combined with other model functions, have been fitted separately to $p_t$ distribution of each measured particle species in a studied collision. However, as indicated in Ref. [34], one cannot assign the physical meaning to collective parameters, such as kinetic freeze-out temperature or radial transverse flow velocity, extracted from model fits to a single particle species. But the simultaneous fits to the spectra of different particle species, extracted in a studied collision type, can result in physically meaningful collective parameters, such as global temperature or/and average transverse flow velocity of a collision system [16,21,33,34]. The combined (global) fits proved to be quite useful and efficient for comparison of the various collision systems using just few parameters [16,21,33,34].

The modern LHC experiments aim to produce the Quark-Gluon Plasma (QGP) in high energy heavy ion-collisions to investigate in detail the properties of QGP matter, which is assumed to have been created a few microseconds after the so called "Big Bang", which is believed to be a starting point for Universe creation. Nevertheless, analysis of small $p+p$ collision systems at high energies is also important and interesting. Such analyses are necessary not only because the results extracted from $p+p$ collisions are regarded as a baseline for analysis of heavy-ion collisions, but also to reveal the collective properties of a system produced in $p+p$ collisions at the highest LHC energies [15]. The reported observations [21, 35-50] of different QGP signatures in high-multiplicity $p+p$ collisions at the LHC, including strong similarity and resemblance of collective features of high-multiplicity $p+p$ collisions to those of heavy ion collisions, serve as a good motivation to investigate further the high energy $p+p$ collisions.

In present work, we analyze the midrapidity $p_t$ distributions of the charged pions and kaons, protons and antiprotons at ten different classes of the average charged-particle (pseudorapidity) multiplicity density ($<dN_{ch}/d\eta>$) in inelastic $p+p$ collisions at $(s)^{1/2}$=7 TeV at the LHC, measured by ALICE collaboration and presented in Ref. [45]. The main goal of present paper is to extract valuable information on evolution of collective properties of a system with changing $<dN_{ch}/d\eta>$ through combined (simultaneous) model fits of the $p_t$ spectra of the charged pions and kaons, protons and antiprotons in each class of charged-particle multiplicity density, using the thermodynamically consistent Tsallis distribution function and Hagedorn formula with the embedded transverse (radial) flow applied over the entire measured long $p_t$ range. The present work follows and uses the same analysis methods as the recent paper [21], in which the midrapidity $p_t$ distributions of identified charged particles at ten different classes of $<dN_{ch}/d\eta>$ in inelastic $p+p$ collisions at $(s)^{1/2}$=13 TeV at the LHC, measured by ALICE collaboration and presented in Ref. [51], have been investigated. This allows one to make a comparative analysis of $p+p$ collision systems at two different energies. Therefore, the results obtained in present work for inelastic $p+p$ collisions at $(s)^{1/2}$=7 TeV are compared systematically with the corresponding results of Ref. [21] extracted for inelastic $p+p$ collisions at $(s)^{1/2}$=13 TeV.

2. **The Data and Models**

In Ref. [45] inelastic $p+p$ collision events at $(s)^{1/2}$=7 TeV with at least one charged particle produced in pseudorapidity range $|\eta|<1$, corresponding to about 75% of the total inelastic scattering cross-section, have been selected by ALICE Collaboration. To study the multiplicity dependence of light-flavor hadron production, the selected collision events have been divided into event classes based on the total charge accumulated in both of V0 detectors (V0M amplitude). The V0M amplitude scales linearly with the total number of the charged particles produced in the pseudorapidity window corresponding to acceptance of V0 scintillators [45]. The corresponding average charged-particle pseudorapidity densities, *<dN$_{ch}$/dη>*, for each class of events have been measured at mid-pseudorapidity ($|\eta|$<0.5). The low $p_t$ part of the particle spectra has been reconstructed for rapidity interval $|y|$<0.5, while the high $p_t$ part has been extracted for the pseudorapidity range $|\eta|$<0.8 in order to take advantage of the full statistics of inelastic *p+p* collisions [45]. The average charged-particle multiplicity density and corresponding fractions of inelastic cross-sections for different classes of event multiplicity are presented in Table 1. The $p_t$ ranges [45] of midrapidity transverse momentum distributions in inelastic *p+p* collisions at $(s)^{1/2}$=7 TeV, measured by ALICE Collaboration, are as follows: [0.1-20.0] GeV/*c* for $\pi^+ + \pi^-$, [0.2-20.0] GeV/*c* for $K^+ + K^-$, and [0.3-20.0] GeV/*c* for $p+\bar{p}$. These $p_t$ ranges coincide with the corresponding $p_t$ ranges for the charged pions and kaons, protons and antiprotons, measured [21] by ALICE Collaboration at midrapidity in ten different classes of charged-particle multiplicity in inelastic *p+p* collisions at $(s)^{1/2}$=13 TeV. It is important to mention that the similar experimental procedures for the event class selection and measurements of corresponding $p_t$ spectra of particles at midrapidity in inelastic *p+p* collisions at $(s)^{1/2}$=7 and 13 TeV have been used by ALICE Collaboration in Refs. [45] and [21], respectively.

**Table 1.** Mean charged-particle multiplicity density in ten different classes of charged-particle multiplicity in inelastic *p+p* collisions at $(s)^{1/2}$=7 TeV.

| V0M mult. class | $< dN_{ch}/d\eta >$ | $\sigma/\sigma_{INEL>0}$ (%) |
|---|---|---|
| I | 21.3±0.6 | 0-0.95 |
| II | 16.5±0.5 | 0.95-4.7 |
| III | 13.5±0.4 | 4.7-9.5 |
| IV | 11.5±0.3 | 9.5-14 |
| V | 10.1±0.3 | 14-19 |
| VI | 8.45±0.25 | 19-28 |
| VII | 6.72±0.21 | 28-38 |
| VIII | 5.40±0.17 | 38-48 |
| IX | 3.90±0.14 | 48-68 |
| X | 2.26±0.12 | 68-100 |

It is widely established that the high $p_t$ part of transverse momentum spectra of hadrons in high-energy nucleon-nucleon collisions can be described well by the Quantum Chromodynamics (QCD) inspired Hagedorn function [52]:

$$\frac{d^2N}{2\pi N_{ev} p_t dp_t dy} = C \left(1 + \frac{m_t}{p_0}\right)^{-n}. \tag{1}$$

Here $C$ is the fitting constant, $p_0$ and $n$ are free parameters, and $m_t = \sqrt{p_t^2 + m_0^2}$ - the transverse mass (energy), $m_0$ is the rest mass of a hadron. As mentioned above, the Tsallis function [1,2] can describe well the invariant $p_t$ and $m_t$ distributions of hadrons in high-energy $p+p$ collisions at the RHIC and LHC [3-10]. There exist several equivalent versions [10-13,22,53] of Tsallis distribution function, which provide practically equal quality fits of transverse momentum distributions of hadrons in high-energy $p+p$ collisions. In its simplest version [10,11,20], the Tsallis function is given at midrapidity ($y \approx 0$) by equation

$$\frac{d^2N}{2\pi N_{ev} p_t dp_t dy} = C_q \left(1 + (q-1)\frac{m_t}{T}\right)^{-1/(q-1)}, \tag{2}$$

where $C_q$ is the fitting constant, $T$ is the effective temperature, and $q$ is the so called non-extensivity parameter, characterizing the degree of deviation of the $p_t$ distribution from the exponential Boltzmann-Gibbs distribution. The function in Eq. (2) is called as a non-extensive generalization of the exponential Boltzmann-Gibbs distribution ($\sim \exp(-\frac{E}{T})$) with the new parameter $q$ added to temperature parameter. The $q$ parameter is also said to be a measure of non-thermalization [54]. When $q$ tends to one (1), the Tsallis function approaches the usual exponential Boltzmann-Gibbs distribution. The closer the parameter $q$ to one, the larger the thermalization degree of a system is.

Compared to various versions of Tsallis distribution functions, the following form [3,9,10] at midrapidity leads to a consistent thermodynamics for the pressure, particle number, and energy density:

$$\frac{d^2N}{2\pi N_{ev} p_t dp_t dy} = C_q m_t \left(1 + (q-1)\frac{m_t}{T}\right)^{-q/(q-1)}, \tag{3}$$

which is called the thermodynamically consistent Tsallis distribution function in present analysis. The values of parameter $T$, extracted from fits by Tsallis functions in Eqs. (2) and (3), represent the effective temperatures, incorporating the contributions of both the thermal motion and collective flow of matter. To separate the thermal motion and collective flow effects, the transverse flow velocity is incorporated into the Tsallis distribution function. It has been demonstrated in Refs. [10,16] that the thermodynamically consistent Tsallis function, given in Eq. (3), results in noticeably smaller values of the effective temperature compared to $T$ values obtained using the simple Tsallis function without thermodynamical consistency in Eq. (2). This has been deduced to be due to the extra $m_t$ term in Eq. (3).

The functions in Eqs. (1) and (2) are mathematically equivalent when one sets $n = 1/(q-1)$ and $p_0 = nT$. The larger values of $n$ correspond to smaller values of $q$. For the quark-quark point scattering $n \approx 4$, and the parameter $n$ becomes larger when multiple scattering centers are involved [6,55,56]. To embed the temperature $T_0$ parameter, the function in Eq. (1) is modified substituting $p_0 = nT_0$:

$$\frac{d^2N}{2\pi N_{ev} p_t dp_t dy} = C\left(1 + \frac{m_t}{n\,T_0}\right)^{-n}. \tag{4}$$

In present work, we incorporate the collective transverse (radial) flow into Eq. (4) by performing the simple transformation $m_t \rightarrow \langle \gamma_t \rangle (m_t - p_t \langle \beta_t \rangle)$, as done also in Refs. [11,15,16,21,33]. Then Eq. (4) modifies to

$$\frac{d^2N}{2\pi N_{ev} p_t dp_t dy} = C\left(1 + \langle \gamma_t \rangle \frac{(m_t - p_t \langle \beta_t \rangle)}{nT_0}\right)^{-n}. \tag{5}$$

Here $<\gamma_t> = 1/\sqrt{1 - <\beta_t>^2}$, $<\beta_t>$ is the average transverse (radial) flow velocity, and $T_0$ - an estimate of kinetic freeze-out temperature. The function in Eq. (5) is called the Hagedorn function (formula) with the embedded transverse flow in present work.

It is important to note that the function in Eq. (5) has been used successfully in Refs. [11,15,16,21,33]. The simple model represented by the function in Eq. (5) is a powerful tool, which probes the long $p_t$ ranges of particles, allowing to compare different collision systems using few parameters [15,16,21,33]. The global parameters $<\beta_t>$ and $T_0$ obtained in Ref. [33] in central copper-copper (Cu+Cu), central gold-gold (Au+Au), and central lead-lead (Pb+Pb) collisions at midrapidity at center-of-mass energy per nucleon pair $(s_{nn})^{1/2}$=200-2760 GeV at the RHIC and LHC, applying the Hagedorn function with the embedded transverse over the long $p_t$ range, have reproduced qualitatively well all the observed dependencies of $<\beta_t>$ and $T_0$ on $<N_{part}>$ and $(s_{nn})^{1/2}$, extracted using three different transverse expansion (blast-wave) models in the low $p_t$ range. The values of the transverse flow velocity, $<\beta_t>$, in the most central (0-5%) Pb+Pb collisions at $(s_{nn})^{1/2}$=2.76 and 5.02 TeV, extracted using Hagedorn function with the embedded transverse flow, applied over long fitting $p_t$ ranges in Ref. [16], have coincided within fit errors with the corresponding values of $\langle \beta_t \rangle$, extracted by ALICE Collaboration in the most central (0-5%) Pb+Pb collisions at $(s_{nn})^{1/2}$=2.76 and 5.02 TeV in Refs. [34] and [57], respectively, using the combined Boltzmann-Gibbs blast-wave fits applied over low $p_t$ ranges. Reproducing satisfactorily the absolute values of average transverse flow velocity, the Hagedorn function with the embedded transverse flow describes qualitatively well the behavior of the centrality dependence of kinetic freeze-out temperatures, extracted from combined Boltzmann-Gibbs blast-wave fits to the particle spectra [16]. Combined fits with Hagedorn function with the embedded transverse flow applied over long $p_t$ ranges in Ref. [16] have confirmed that $\langle \beta_t \rangle$ increases and $T_0$ decreases with an increase in collision centrality in Pb+Pb collisions at $(s_{nn})^{1/2}$=2.76 and 5.02 TeV, which is in very good agreement with the similar result of the ALICE Collaboration obtained from combined Boltzmann-Gibbs blast-wave fits to the particle spectra in Pb+Pb collisions at $(s_{nn})^{1/2}$=2.76 and 5.02 TeV in the low $p_t$ ranges in Refs. [34,57].

In present analysis, for description of the $p_t$ distributions, $d^2N/(N_{ev}dp_tdy)$, of the identified charged particles in different event multiplicity classes in inelastic $p+p$ collisions at $(s)^{1/2}$=7 TeV, measured by ALICE Collaboration [45], we modify the thermodynamically consistent Tsallis function in Eq. (3) as follows:

$$\frac{d^2N}{N_{ev}dp_tdy} = 2\pi C_q p_t\, m_t \left(1 + (q-1)\frac{m_t}{T}\right)^{-\frac{q}{q-1}}, \qquad (6)$$

Analogously, for describing the $p_t$ spectra, $d^2N/(N_{ev}dp_tdy)$, of particles in present work, the Hagedorn function with the embedded transverse flow in Eq. (5) is modified to expression

$$\frac{d^2N}{N_{ev}dp_tdy} = 2\pi C\, p_t \left(1 + \langle \gamma_t \rangle \frac{(m_t - p_t \langle \beta_t \rangle)}{nT_0}\right)^{-n}. \qquad (7)$$

In present analysis, the combined (simultaneous) fits with the theoretical model functions, given in Eqs. (6) and (7), of the $p_t$ spectra of particle species in each class of event multiplicity have been performed using the Nonlinear Curve Fitting of the Origin 9.1 Data Analysis and Graphing Software. The $T/T_0$ or/and $\langle \beta_t \rangle$ parameters have been shared (global) parameters for the analyzed particle species during the fit procedures. The combined statistical and systematic errors (added in quadrature) are given for the experimental data points in the figures of present paper. The procedures for calculation of systematic errors in $p_t$ spectra of the analyzed charged particles are given in Ref. [45]. The minimum $\chi^2$ fit procedures have been performed taking into account the combined statistical and systematic errors as the weights (1/(error)$^2$) for the data points. While fitting the $p_t$ spectra of charged pions, the region $p_t$ < 0.5 GeV/$c$, containing significant contribution from decays of baryon resonance, has been excluded, as done also in Refs. [15,16,21,33,34,57]. It is important to mention that in present work we use the same model functions, given in Eqs. (6) and (7), and identical fitting $p_t$ ranges for particle species as used in recent work [21] for inelastic $p+p$ collisions at $(s)^{1/2}$=13 TeV.

## 3. Analysis and Results

The results extracted from simultaneous minimum $\chi^2$ fits with the thermodynamically consistent Tsallis function (Eq. (6)) of the $p_t$ distributions of identified charged particles in different classes of charged-particle multiplicity in inelastic $p+p$ collisions at $(s)^{1/2}$=7 TeV are presented in Table 2. As an example, the corresponding combined minimum $\chi^2$ fit curves by the thermodynamically consistent Tsallis function of the experimental midrapidity $p_t$ distributions of the charged pions and kaons, protons and antiprotons in four different classes of charged-particle multiplicity density are demonstrated in Fig. 1.

**Table 2.** Parameters extracted from combined minimum $\chi^2$ fits with the thermodynamically consistent Tsallis function (Eq. (6)) of the $p_t$ spectra of identified charged particles in different classes of charged-particle multiplicity density in inelastic $p+p$ collisions at $(s)^{1/2}$=7 TeV. The fitted $p_t$ ranges are as follows: [0.5-20.0] GeV/$c$ for $\pi^++\pi^-$, [0.2-20.0] GeV/$c$ for $K^++K^-$, and [0.3-20.0] GeV/$c$ for $p+\bar{p}$. Here $n.d.f.$ denotes the number of degrees of freedom.

| $<dN_{ch}/d\eta>$ | $q$ ($\pi^++\pi^-$) | $q$ ($K^++K^-$) | $q$ ($p+\bar{p}$) | $T$ (MeV) | $\chi^2/n.d.f.$ ($n.d.f.$) |
|---|---|---|---|---|---|
| 21.3±0.6 | 1.143±0.001 | 1.153±0.001 | 1.133±0.001 | 131±2 | 1.85 (114) |
| 16.5±0.5 | 1.145±0.001 | 1.154±0.001 | 1.131±0.001 | 121±2 | 1.54 (114) |
| 13.5±0.4 | 1.146±0.001 | 1.155±0.001 | 1.130±0.001 | 114±1 | 1.06 (114) |
| 11.5±0.3 | 1.147±0.001 | 1.155±0.001 | 1.128±0.001 | 109±1 | 0.79 (114) |
| 10.1±0.3 | 1.148±0.001 | 1.155±0.001 | 1.128±0.001 | 104±1 | 0.65 (114) |
| 8.45±0.25 | 1.149±0.001 | 1.156±0.001 | 1.128±0.001 | 98±1 | 0.61 (114) |
| 6.72±0.21 | 1.150±0.001 | 1.156±0.001 | 1.125±0.001 | 91±1 | 0.45 (114) |
| 5.40±0.17 | 1.151±0.001 | 1.157±0.001 | 1.123±0.001 | 85±1 | 0.39 (114) |
| 3.90±0.14 | 1.151±0.001 | 1.156±0.001 | 1.122±0.001 | 75±1 | 0.79 (114) |
| 2.26±0.12 | 1.149±0.001 | 1.153±0.001 | 1.114±0.001 | 57±1 | 1.00 (114) |

**Table 3.** Parameters extracted from combined minimum $\chi^2$ fits with the Hagedorn formula with the embedded transverse flow (Eq. (7)) of the $p_t$ spectra of identified charged particles in different classes of charged-particle multiplicity density in inelastic $p+p$ collisions at $(s)^{1/2}$=7 TeV. The fitted $p_t$ ranges are as follows: [0.5-20.0] GeV/$c$ for $\pi^++\pi^-$, [0.2-20.0] GeV/$c$ for $K^++K^-$, and [0.3-20.0] GeV/$c$ for $p+\bar{p}$.

| $<dN_{ch}/d\eta>$ | $n$ ($\pi^++\pi^-$) | $n$ ($K^++K^-$) | $n$ ($p+\bar{p}$) | $\langle\beta_t\rangle$ (in $c$ units) | $T_0$ (MeV) | $\chi^2/n.d.f.$ ($n.d.f.$) |
|---|---|---|---|---|---|---|
| 21.3±0.6 | 6.80±0.05 | 6.61±0.04 | 8.11±0.06 | 0.29±0.02 | 113±3 | 0.60 (113) |
| 16.5±0.5 | 6.77±0.05 | 6.57±0.04 | 8.13±0.06 | 0.23±0.02 | 113±3 | 0.72 (113) |
| 13.5±0.4 | 6.78±0.04 | 6.55±0.04 | 8.17±0.06 | 0.18±0.02 | 114±3 | 0.63 (113) |
| 11.5±0.3 | 6.76±0.04 | 6.52±0.04 | 8.24±0.06 | 0.15±0.02 | 113±3 | 0.54 (113) |
| 10.1±0.3 | 6.77±0.04 | 6.52±0.04 | 8.24±0.06 | 0.12±0.02 | 115±3 | 0.58 (113) |
| 8.45±0.25 | 6.74±0.04 | 6.48±0.04 | 8.15±0.06 | 0.07±0.02 | 115±4 | 0.74 (113) |
| 6.72±0.21 | 6.74±0.04 | 6.47±0.04 | 8.29±0.06 | 0.04±0.02 | 112±3 | 0.60 (113) |
| 5.40±0.17 | 6.73±0.04 | 6.46±0.03 | 8.40±0.06 | 0±0.02 | 108±3 | 0.56 (113) |
| 3.90±0.14 | 6.70±0.05 | 6.49±0.04 | 8.48±0.09 | 0±0.03 | 95±4 | 0.97 (113) |
| 2.26±0.12 | 6.75±0.06 | 6.59±0.05 | 9.06±0.13 | 0±0.03 | 69±4 | 1.19 (113) |

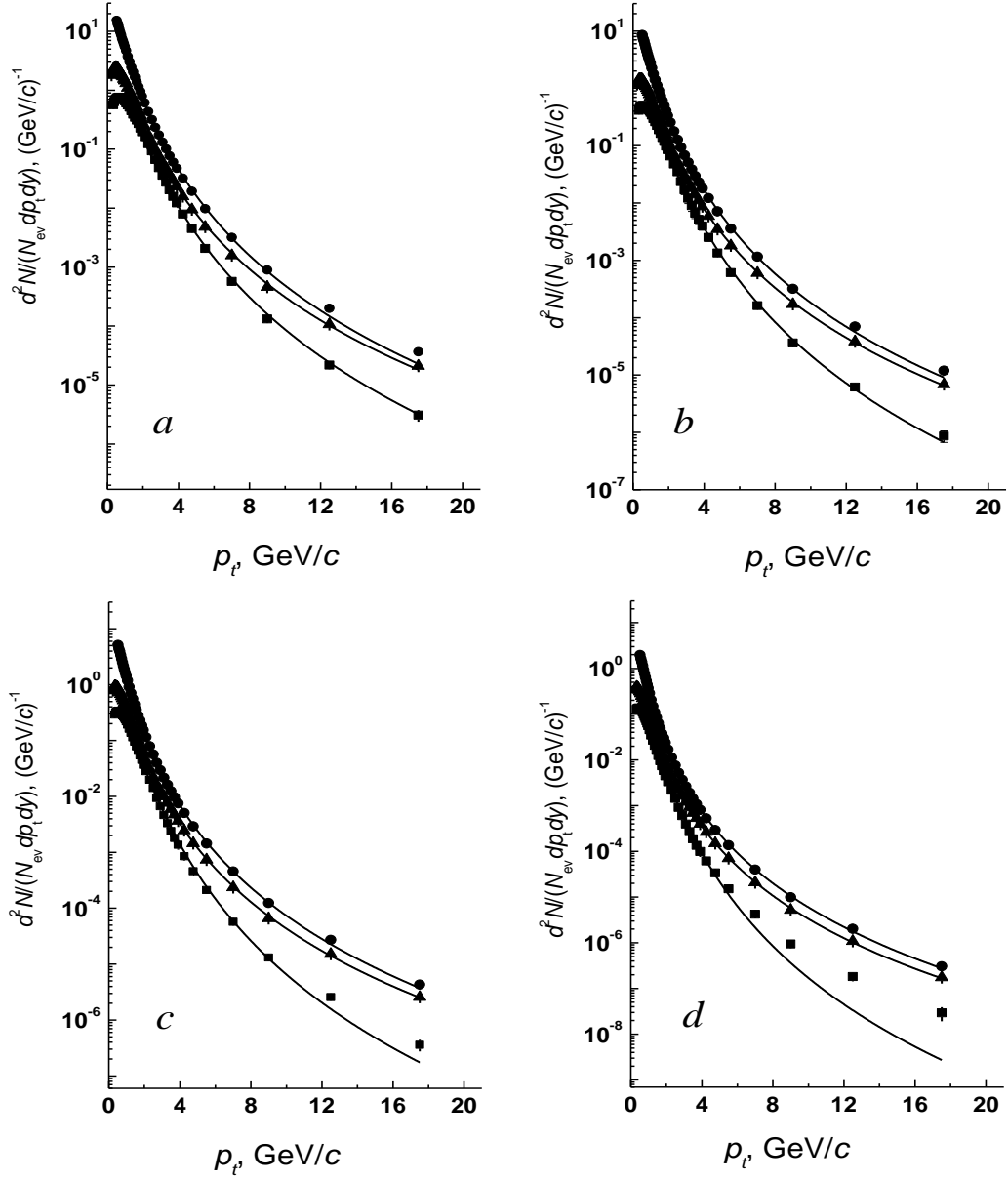

**Figure 1.** The combined (simultaneous) minimum χ2 fits (solid curves) by the function in Eq. (6) of the experimental midrapidity $p_t$ spectra of the charged pions (●) and kaons (▲), protons and antiprotons (■) in inelastic $p+p$ collisions at $(s)^{1/2}=7$ TeV in different classes of charged-particle multiplicity density: $<dN_{ch}/d\eta>$ = 21.3±0.6 (*a*), $<dN_{ch}/d\eta>$ = 11.5±0.3 (*b*), $<dN_{ch}/d\eta>$ = 6.72±0.21 (*c*), and $<dN_{ch}/d\eta>$ = 2.26±0.12 (*d*). The vertical error bars are the quadratic sum of statistical and systematic uncertainties, dominated by the systematic ones.

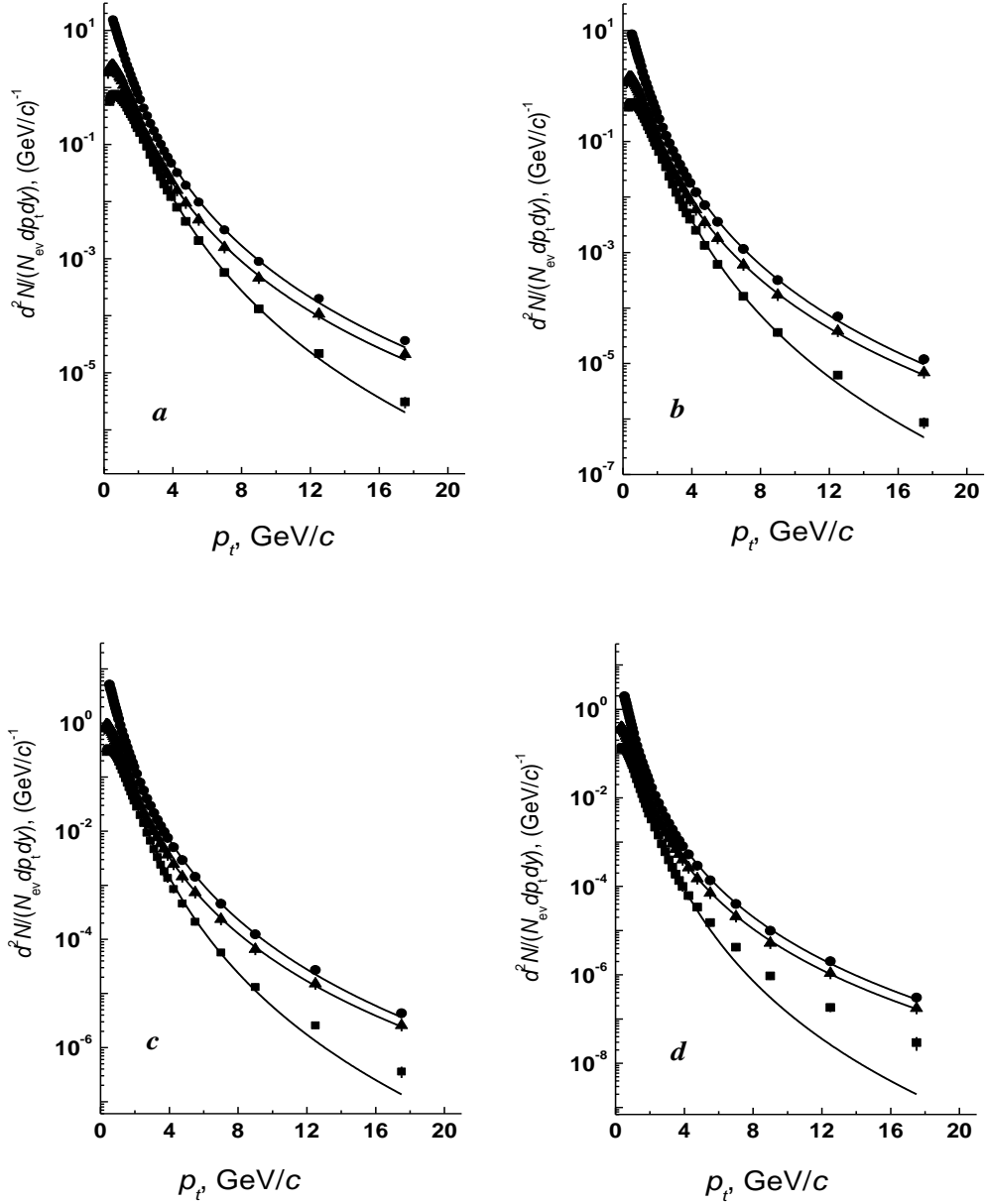

**Figure 2.** The combined (simultaneous) minimum χ2 fits (solid curves) by the function in Eq. (7) of the experimental midrapidity $p_t$ spectra of the charged pions (●) and kaons (▲), protons and antiprotons (■) in inelastic $p+p$ collisions at $(s)^{1/2}$=7 TeV in different classes of charged-particle multiplicity density: $<dN_{ch}/d\eta>$ = 21.3±0.6 (*a*), $<dN_{ch}/d\eta>$ = 11.5±0.3 (*b*), $<dN_{ch}/d\eta>$ = 6.72±0.21 (*c*), and $<dN_{ch}/d\eta>$ = 2.26±0.12 (*d*). The vertical error bars are the quadratic sum of statistical and systematic uncertainties, dominated by the systematic ones.

Parameters obtained from combined minimum $\chi^2$ fits by the Hagedorn function with the embedded transverse flow (Eq. 7) of the transverse momentum spectra of particles in different classes of charged-particle multiplicity density in inelastic $p+p$ collisions at $(s)^{1/2}$=7 TeV are shown in Table 3.

As an example, the corresponding simultaneous minimum $\chi^2$ fit curves by the Hagedorn function with the embedded transverse flow of the experimental midrapidity $p_t$ distributions of the charged pions and kaons, protons and antiprotons in four different classes of charged-particle multiplicity density are illustrated in Fig. 2.

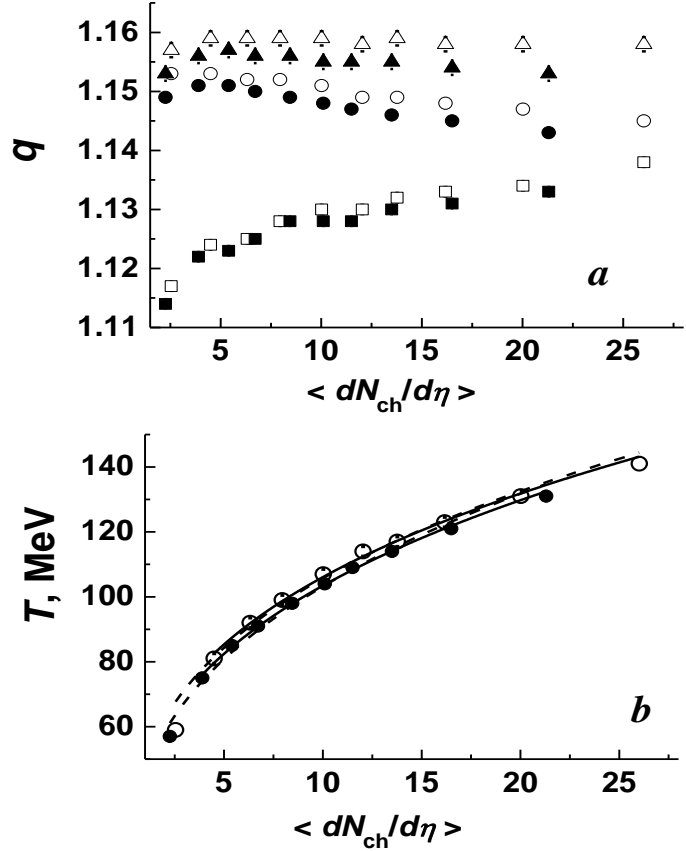

**Figure 3.** (*a*) - The charged-particle multiplicity density $<dN_{ch}/d\eta>$ dependencies of the extracted $q$ values of thermodynamically consistent Tsallis function, presented in Table 2, for the charged pions (●) and kaons (▲), protons and antiprotons (■) in inelastic $p+p$ collisions at $(s)^{1/2}$=7 TeV; (*b*) – the same for the obtained effective temperatures $T$ (●) of thermodynamically consistent Tsallis function, given in Table 2. The corresponding data extracted in Ref. [21] for the charged pions and kaons, protons and antiprotons in inelastic $p+p$ collisions at $(s)^{1/2}$=13 TeV are given by the corresponding open symbols for comparison. The data points have been fitted with the simple power function $T = A \cdot <\frac{dN_{ch}}{d\eta}>^\alpha$, where $A$ is the fitting constant, and $\alpha$ - the exponent parameter. The dashed and solid curves are the simple power function fits of the whole range (10 data points), and of the whole range excluding the first data point (9 data points), respectively.

As seen from Fig. 1 and $\chi^2/n.d.f.$ values in Table 2, the combined fits with thermodynamically consistent Tsallis function reproduce quite satisfactorily the transverse momentum distributions of particles in ten different classes of charged-particle multiplicity in inelastic $p+p$ collisions at $(s)^{1/2}$=7 TeV. Similarly, as observed from Fig. 2 and $\chi^2/n.d.f.$ values in Table 3, the simultaneous fits by the Hagedorn function with the embedded transverse flow describe quite well the $p_t$ distributions of identified charged particles in the analyzed ten classes of charged-particle multiplicity.

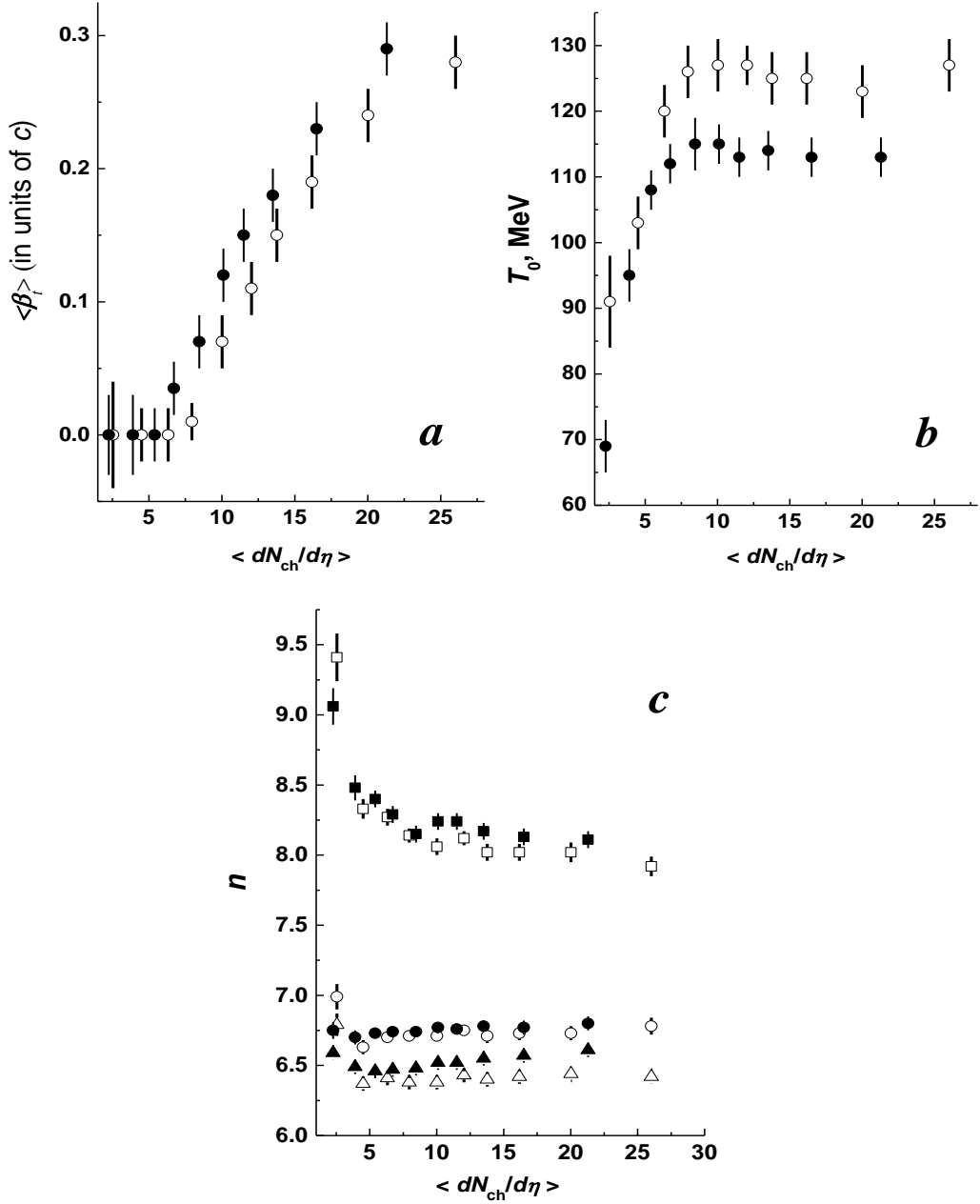

**Figure 4.** The charged-particle multiplicity density $<dN_{ch}/d\eta>$ dependencies of the extracted $\langle \beta_t \rangle$ (*a*) and $T_0$ (*b*) parameters (●) of Hagedorn function with the embedded transverse flow, presented in Table 3, in inelastic $p+p$ collisions at $(s)^{1/2}=7$ TeV; (*c*) – the same for the obtained *n* values of Hagedorn function with the embedded transverse flow, given in Table 3, for the charged pions (●) and kaons (▲), protons and antiprotons (■). The corresponding data extracted in Ref. [21] for the charged pions and kaons, protons and antiprotons in inelastic $p+p$ collisions at $(s)^{1/2}=13$ TeV are given by the corresponding open symbols for comparison.

Figure 3 illustrates the charged-particle multiplicity density $<dN_{ch}/d\eta>$ dependencies of the extracted *q* and *T* parameters of thermodynamically consistent Tsallis function for the charged pions and kaons, protons and antiprotons in inelastic *p+p* collisions at $(s)^{1/2}$=7 TeV, extracted from the combined minimum $\chi^2$ fits and given in Table 2. Figure 4 summarizes the $<dN_{ch}/d\eta>$ dependencies of the obtained $\langle\beta_t\rangle$, $T_0$ and *n* parameters of Hagedorn function with the embedded transverse flow for the charged pions and kaons, protons and antiprotons in inelastic *p+p* collisions at $(s)^{1/2}$=7 TeV, obtained from the combined minimum $\chi^2$ fits and presented in Table 3. The corresponding data obtained in Ref. [21] for the charged pions and kaons, protons and antiprotons in inelastic *p+p* collisions at $(s)^{1/2}$=13 TeV are also presented in Figs. 3 and 4 for comparison. Here it is worth mentioning the importance of analyzing the behavior of $\langle\beta_t\rangle$ and $T_0$ and their excitation functions [14] due to their relation to map the QCD phase diagram, even though the chemical freeze-out temperature ($T_{ch}$) is normally used in such phase diagrams.

**Table 4.** Parameters obtained from minimum $\chi^2$ fits with the simple power function $T = A \cdot <\frac{dN_{ch}}{d\eta}>^\alpha$ of the ***T*** versus $<dN_{ch}/d\eta>$ dependencies in Fig. 3(*b*) in the whole $<dN_{ch}/d\eta>$ range (**I**), and in the whole range excluding the first data point (**II**) with $<dN_{ch}/d\eta>$ = 2.26±0.12 and $<dN_{ch}/d\eta>$ = 2.55±0.04 in *p+p* collisions at $(s)^{1/2}$=7 TeV and 13 TeV, respectively. The corresponding data extracted in Ref. [21] for inelastic *p+p* collisions at $(s)^{1/2}$=13 TeV are shown for comparison.

| $<dN_{ch}/d\eta>$ fit range | Collision type, $(s)^{1/2}$ | *A* (MeV) | $\alpha$ | $\chi^2$/n.d.f. (n.d.f.) |
|---|---|---|---|---|
| **I** | *p+p*, 7 TeV | 45.8±1.4 | 0.353±0.013 | 3.67 (8) |
|  | *p+p*, 13 TeV [21] | 49.9±1.7 | 0.33±0.01 | 3.66 (8) |
| **II** | *p+p*, 7 TeV | 48.7±0.6 | 0.327±0.005 | 0.37 (7) |
|  | *p+p*, 13 TeV [21] | 51.5±0.9 | 0.31±0.01 | 1.05 (7) |

As observed from Fig. 3(*a*), a clear separation of non-extensivity parameter *q* values for mesons (pions and kaons) and baryons (protons and antiprotons) with the relation *q*(baryons) < *q*(mesons) is seen in the whole studied $<dN_{ch}/d\eta>$ range in inelastic *p+p* collisions at $(s)^{1/2}$=7 and 13 TeV. This is in agreement with the relation *q*(baryons) < *q*(mesons) in minimum bias inelastic *p+p* collisions (that is, in the total ensemble of inelastic *p+p* collisions) at high energies obtained earlier in Refs. [15, 16, 22, 25]. As seen from Fig. 3(*a*), the behavior of *q* versus $<dN_{ch}/d\eta>$ dependence for all studied particle species in *p+p* collisions at $(s)^{1/2}$=7 TeV is similar to that obtained in p+p collisions at $(s)^{1/2}$=13 TeV in Ref. [21]. However, as seen from Fig. 3(*a*), on the whole the *q* values for the charged pions and kaons, protons and antiprotons in *p+p* collisions at $(s)^{1/2}$=7 TeV proved to be noticeably smaller as compared to those in *p+p* collisions at $(s)^{1/2}$=13 TeV in the whole analyzed *<dN$_{ch}$/dη>* range. This difference is pronounced more for the charged pions and kaons as compared to the protons and antiprotons. This can suggest that the systems produced in *p+p* collisions at $(s)^{1/2}$=7 TeV are characterized by the larger degree of equilibrium and thermalization as compared to that in *p+p* collisions at $(s)^{1/2}$=13 TeV. This result is consistent with that of the recent work [16], in which non-extensivity parameter *q* has been shown to increase systematically for all the studied particle species with increasing the energy $(s)^{1/2}$ of *p+p* collisions from 2.76 to 5.02 TeV, implying that the faster and more violent *p+p* collisions at $(s)^{1/2}$= 5.02 TeV result in a smaller degree of thermalization (larger degree of non-equilibrium) compared to that in *p+p* collisions at $(s)^{1/2}$=2.76 TeV [16].

In Ref. [8] the combined $p_t$ distributions of the charged particles, consisting predominantly of pions, in minimum bias *p+p* collisions at wide energy range $(s)^{1/2}$=0.54-7 TeV have been analyzed using the thermodynamically consistent Tsallis distribution. It has been obtained [8] that the non-extensivity parameter *q* increases clearly with beam energy reaching the highest value ≈1.15 in minimum bias *p+p* collisions at $(s)^{1/2}$=7 TeV. As seen from Table 2, this value agrees well with *q*≈1.15 for charged pions, extracted in present analysis, in majority of multiplicity classes in *p+p* collisions at $(s)^{1/2}$=7 TeV. The effective Tsallis temperature 82±1 MeV obtained in Ref. [8] in minimum bias *p+p* collisions at $(s)^{1/2}$=7 TeV has coincided within uncertainties with the average effective Tsallis temperature *<T>*=81±1 MeV calculated in present analysis. In present work, the average effective Tsallis temperature *<T>* for minimum bias *p+p* collisions at $(s)^{1/2}$=7 TeV has been evaluated using the extracted temperature $T_i$ from Table 2 and corresponding fraction of events $f_i$ (with respect to the total inelastic cross-section) of *i-th* class of charged-particle multiplicity density from Table 1 as follows: $<T> = \sum_{i=1}^{10}(T_i \cdot f_i)$.

As seen from Fig. 3(*b*), the obtained effective temperatures *T* of thermodynamically consistent Tsallis function show a similar consistent increase with increasing the multiplicity of charged particles in inelastic *p+p* collisions at $(s)^{1/2}$=7 TeV as that observed in inelastic *p+p* collisions at $(s)^{1/2}$=13 TeV. This is quite natural result as the larger multiplicity events correspond to harder *p+p* collisions with higher energy-momentum transferred to a system [21]. The data points in Fig. 3(*b*) extracted in *p+p* collisions at $(s)^{1/2}$=7 TeV have been fitted with the simple power function $T = A \cdot <\frac{dN_{ch}}{d\eta}>^\alpha$, where *A* is the fitting constant, and *α* - the exponent parameter, as done in Ref. [21]. Similarly to Ref. [21], the minimum $\chi^2$ fits by this simple power function have been performed in the whole analyzed *<dN$_{ch}$/dη>* range, and in the whole range excluding the first data point, that is, excluding the softest $p+p$ collisions. The corresponding minimum $\chi^2$ fit results are presented in Table 4. The corresponding data extracted in Ref. [21] for inelastic $p+p$ collisions at $(s)^{1/2}$=13 TeV are also shown in Table 4 for comparison. As seen from Fig. 3(*b*) and $\chi^2$/*n.d.f.* values in Table 4, the simple power function cannot fit satisfactorily the *T* versus *<dN$_{ch}$/dη>* dependence in the whole range. On the other hand, as observed from Fig. 3(*b*) and $\chi^2$/*n.d.f.* values in Table 4, the *T* versus *<dN$_{ch}$/dη>* dependence in $p+p$ collisions at $(s)^{1/2}$=7 TeV is described quite satisfactorily with this simple power function with exponent parameter ≈1/3 in region *<dN$_{ch}$/dη>* > 3, when the first data point is excluded from the fit range. This agrees very well, as seen from Fig. 3(*b*) and Table 4, with the similar result of Ref. [21], in which the *T* versus *<dN$_{ch}$/dη>* dependence in $p+p$ collisions at $(s)^{1/2}$=13 TeV has been described quite well with the simple power function $T = A \cdot <\frac{dN_{ch}}{d\eta}>^\alpha$ with exponent parameter ≈1/3 in the whole *<dN$_{ch}$/dη>* range, excluding the first data point.

As seen from Fig. 4(*a*), the average value of transverse flow velocity, extracted using Hagedorn function with the embedded transverse flow, proved to be essentially zero in region *<dN$_{ch}$/dη>* < 6 of low multiplicity density (the first three data points) in $p+p$ collisions at $(s)^{1/2}$=7 TeV. As observed from Fig. 4(*a*), beginning from approximately *<dN$_{ch}$/dη>* ≈ 6, the transverse flow starts emerging and developing with parameter $\langle \beta_t \rangle$ growing steadily in region *<dN$_{ch}$/dη>* > 6 up to the highest analyzed values of the average multiplicity density. The related parameter $T_0$ (which approximates the kinetic freeze-out temperature) grows consistently in region *<dN$_{ch}$/dη>* < 6 of low multiplicity density (the first three data points) and remains constant within the fit errors in the wide range *<dN$_{ch}$/dη>* > 6 in $p+p$ collisions at $(s)^{1/2}$=7 TeV. The observed broad plateau region of $T_0$ (starting roughly at *<dN$_{ch}$/dη>* ≈ 6) in Fig. 4(*b*) corresponds to simultaneous and consistent growth of $\langle \beta_t \rangle$ with an increase in multiplicity of charged particles in the wide studied range *<dN$_{ch}$/dη>* > 6 in Fig. 4(*a*). The fact that the temperature $T_0$ first grows in region *<dN$_{ch}$/dη>* < 6, reaching then a broad plateau starting at *<dN$_{ch}$/dη>* ≈ 6, which matches the practical absence of transverse flow at *<dN$_{ch}$/dη>* < 6 and development and consistent rise of $\langle \beta_t \rangle$ at wide range *<dN$_{ch}$/dη>* > 6, indicates the probable deconfinement phase transition at around *<dN$_{ch}$/dη>* ≈ 6 in inelastic $p+p$ collisions at $(s)^{1/2}$=7 TeV. The $\langle \beta_t \rangle$ and $T_0$ versus *<dN$_{ch}$/dη>* dependencies in $p+p$ collisions at $(s)^{1/2}$=7 TeV, as seen from Figs. 4(*a*) and 4(*b*), match qualitatively those in $p+p$ collisions at $(s)^{1/2}$=13 TeV obtained in Ref. [21] with some differences in the absolute values of parameters. It is seen from Figs. 4(*a*) and 4(*b*) that the absolute values of $\langle \beta_t \rangle$ are noticeably smaller and *<dN$_{ch}$/dη>* value for probable phase transition and value of $T_0$ in plateau region are larger in $p+p$ collisions at $(s)^{1/2}$=13 TeV as compared to those in $p+p$ collisions at $(s)^{1/2}$=7 TeV. It has been estimated in Ref. [21] from analysis of $T_0$ and $\langle\beta_t\rangle$ versus $<dN_{ch}/d\eta>$ dependencies that the probable deconfinement phase transition in inelastic $p+p$ collisions at $(s)^{1/2}$=13 TeV occurs at $<dN_{ch}/d\eta> \approx 7.1\pm0.2$. We have followed the similar procedure as in Ref. [21] to better estimate the critical value of $<dN_{ch}/d\eta>$ for probable deconfinement phase transition in $p+p$ collisions at $(s)^{1/2}$=7 TeV: the critical $<dN_{ch}/d\eta>$ is estimated to be as the middle value on $<dN_{ch}/d\eta>$ axis between the 3$^{rd}$ and 4$^{th}$ points in Fig. 4(*b*), corresponding to $<dN_{ch}/d\eta>$=5.40±0.17 and $<dN_{ch}/d\eta>$=6.72±0.21, respectively (see Table 1). The corresponding estimate for probable deconfinement phase transition in $p+p$ collisions at $(s)^{1/2}$=7 TeV proved to be $<dN_{ch}/d\eta> \approx 6.1\pm0.3$, which is noticeably lower than corresponding $<dN_{ch}/d\eta> \approx 7.1\pm0.2$ obtained in $p+p$ collisions at $(s)^{1/2}$=13 TeV in Ref. [21].

As seen from Fig. 4(*c*), the observed *n* versus $<dN_{ch}/d\eta>$ dependencies, extracted using Hagedorn function with the embedded transverse flow, in $p+p$ collisions at $(s)^{1/2}$=7 and 13 TeV [21] match qualitatively the corresponding inverse dependencies of the *q* parameter for the corresponding particle species, observed in Fig. 3(*a*).

## 4. Analysis and Discussion of Results

It has been observed that the transverse (radial) flow becomes significant at higher multiplicity events in $p+p$ collisions at $(s)^{1/2}$=7 TeV, attaining the maximum value $\langle\beta_t\rangle$ = 0.29±0.02 (see Table 3) at the largest average multiplicity density $<dN_{ch}/d\eta>$=21.3±0.6. At the same time, the transverse flow has been absent in low multiplicity $p+p$ collisions in region $<dN_{ch}/d\eta>$ < 6 (see Table 3 and Fig. 4(*a*)). These results are consistent with the similar emergence and development of transverse radial flow in higher charged-particle multiplicity events in $p+p$ collisions at $(s)^{1/2}$=13 TeV at the LHC obtained in Ref. [21]. As seen from Table 1, the zero transverse flow collisions with $<dN_{ch}/d\eta>$ < 6 make up around 62% of the total ensemble of inelastic $p+p$ collisions at $(s)^{1/2}$=7 TeV. In recent works [15,16] the negligible transverse flow velocity, consistent with zero value within the fit uncertainties, has been extracted in minimum bias inelastic $p+p$ collisions at $(s)^{1/2}$=2.76, 5.02, and 7 TeV analyzing the midrapidity $p_t$ spectra of identified charged particles with the Hagedorn function with the embedded transverse flow. Practically zero transverse flow velocity obtained in Refs. [15,16] in minimum bias inelastic $p+p$ collisions at $(s)^{1/2}$=2.76, 5.02, and 7 TeV is likely due to the fact that the dominant fraction of analyzed inelastic $p+p$ events are low multiplicity events with practical absence of transverse flow.

The different QGP signatures: the strangeness enhancement [38], hardening of transverse momentum spectra [39,40], and other QGP-like characteristics [35-37, 41-45, 46-50,58] obtained in $p+p$ collisions at the LHC have been reported. The collective expansion by emerging hadrons and onset of collective radial flow in $p+p$ collisions at $(s)^{1/2}$=900 and 7000 GeV at the LHC has been demonstrated in Ref. [35]. The emergence and development of transverse radial flow in higher charged-particle multiplicity events in $p+p$ collisions at $(s)^{1/2}$=13 TeV at the LHC have been reported in Ref. [21]. The signatures of an equilibrated and collective system in high multiplicity $p+p$ collisions at $(s)^{1/2}$=7 TeV at the LHC have been found in Ref. [45]. In Ref. [38] the integrated yields of strange and multi-strange particles, relative to pions, have been shown to increase significantly with increasing the event charged-particle multiplicity in $p+p$ collisions at $(s)^{1/2}$=7 TeV at the LHC. This has been the first reported observation [38] of strangeness enhancement in high-multiplicity $p+p$ collisions. These measurements [38] of ALICE Collaboration proved to be in remarkable agreement with those of $p+Pb$ collisions at $(s_{nn})^{1/2}$=5.02 TeV at the LHC, implying that the observed phenomenon is due to the final system created in a collision. In high-multiplicity $p+p$ collisions [38] the strangeness production has reached values similar to those obtained in Pb+Pb collisions at $(s_{nn})^{1/2}$=2.76 TeV, in which QGP is produced. In recent analysis [44], the transverse momentum distributions of identified particles as a function of multiplicity of charged particles and transverse spherocity have been investigated with the help of non-extensive Tsallis statistics and Boltzmann-Gibbs Blastwave (BGBW) model in $p+p$ collisions at $(s)^{1/2}$=13 TeV using PYTHIA8 event generator. It has been deduced that the isotropic $p+p$ events approach to thermal equilibrium, while the jetty ones remain far from equilibrium [44].

Nowadays [58] multiparticle production in both $p+p$ and nucleus-nucleus ($A+A$) collisions is described via color strings stretched between the projectile and target. They (strings) decay into new strings and subsequently hadronize, producing the observed hadrons [58]. With growth of energy and size of the colliding system, the number of strings increase, and they start to overlap, producing clusters in the transverse plane.

The recent analyses made in Refs. [46-50] explain quite convincingly the origin of thermalization in high-energy $p+p$ collisions. The exponential component in $p_t$ spectra of particles with approximate thermal abundances of the hadron yields in high-energy $p+p$ collisions are considered as a clear sign of thermalization [46,50,59]. Because few secondary interactions in high-energy $p+p$ collisions do not favor thermalization through conventional final-state interactions, an occurrence of thermal feature in $p+p$ collisions has been quite surprising [46,50]. The thermalization in high-energy $p+p$ collisions has been explained in Refs. [46-49] as that occurring during the rapid quench induced by the collisions due to the high degree of quantum entanglement inside the partonic wave functions of colliding protons. Therefore, the effective temperature obtained from $p_t$ spectra of particles depends on the momentum transfer, representing an ultraviolet cutoff of quantum modes resolved by a collision [46, 50]. In Ref. [50] the transverse momentum spectra at various multiplicities of both Pb+Pb collisions at $(s_{nn})^{1/2}$=2.76 TeV and p+p interactions at $(s)^{1/2}$=7 TeV have been fitted well by sum of an exponential plus power like function, represented by thermal-like temperature, $T_{th}$, and hard temperature, $T_h$, respectively. The thermalization caused by quantum entanglement has been confirmed in Ref. [50] from analysis of multiplicity dependence of the p+p and Pb+Pb data at the LHC. No thermal radiation is expected in diffractive p+p events with a large rapidity gap, because in such events the whole wave function of the proton gets involved and no entanglement entropy occurs [50]. It has been indeed observed in Ref. [46] that the thermal component vanishes in diffractive p+p events at $(s)^{1/2}$=13 TeV in spite of many hadrons being still produced. In the studied p+p and Pb+Pb data at the LHC, the effective thermalization temperature, $T_{th}$, has been deduced [50] to be proportional to the hard temperature, $T_h$, defined by the average transverse momentum. The obtained proportionality coefficient has been shown to be universal and independent of the collision type [50]. The proportionality between $T_{th}$ and $T_h$ has been explained naturally as that arising from the clustering of color sources [50]. The thermal and hard temperatures have been found to increase with increasing multiplicity in both p+p and Pb+Pb collisions at the LHC [50].

The consistent increase of the effective temperature $T$ (extracted using thermodynamically consistent Tsallis function) with increasing multiplicity of charged particles in inelastic p+p collisions at $(s)^{1/2}$=7 and 13 TeV [21], as seen from Fig. 3(*b*), is in agreement with the increase with multiplicity of thermal temperature, $T_{th}$, in p+p collisions at the LHC in Ref. [50]. This could be explained by the fact that the bigger multiplicity events correspond to harder p+p collisions with larger amount of energy-momentum transferred [21]. In present paper, we have confirmed an interesting result of Ref. [21] that the $T$ versus <$dN_{ch}/d\eta$> dependence in p+p collisions at $(s)^{1/2}$=13 TeV is described very well by the simple power function $T = A \cdot <\frac{dN_{ch}}{d\eta}>^\alpha$ with exponent parameter $\approx 1/3$. We have obtained in present analysis that the $T$ versus <$dN_{ch}/d\eta$> dependence in p+p collisions at $(s)^{1/2}$=7 TeV is also reproduced quite satisfactorily with the simple power function $T = A \cdot <\frac{dN_{ch}}{d\eta}>^\alpha$ with exponent parameter $\approx 1/3$ in the whole analyzed range <$dN_{ch}/d\eta$> > 3. Hence, assuming the proportionality of the initial energy density ($\varepsilon$) to the charged-particle multiplicity density, it is obtained that $T \sim \varepsilon^{1/3}$ in inelastic p+p collisions at $(s)^{1/2}$=7 and 13 TeV. Similarly to Ref. [21], we can compare this result with the known relation $u = A \cdot T^4$ ($A$ – some constant of integration) between the energy density, *u*, of blackbody radiation and its thermodynamic temperature, *T*, regarded as one form of the Stefan-Boltzmann law. Comparing $T \sim \varepsilon^{1/3}$ extracted in present analysis and in Ref. [21] with relation $T \sim u^{1/4}$ for blackbody radiation, it is seen that the energy density dependencies of the effective temperature of the systems, created in *p+p* collisions at $(s)^{1/2}$=7 and 13 TeV [21], and of the blackbody temperature are close to each other due to closeness of the respective exponent parameters. It is interesting to mention that in a simple model of an ideal gas of massless pions [60] the energy density as a function of temperature is also described by the Stefan-Boltzmann form $\varepsilon_\pi = \left(\frac{\pi^2}{10}\right) T^4$, and hence $T \sim \varepsilon_\pi^{1/4}$. The difference between exponent parameter (1/3) obtained in present work and that (1/4) for an ideal gas of massless pions could probably be explained by nonzero viscosity to entropy ratio (($\eta/s$) > 0) and some nonzero coupling between the constituents of the dense medium produced in high-energy proton-proton collisions at the LHC.

Using the exact and explicit solutions of relativistic hydrodynamics, the advanced estimates of initial energy density in minimum bias *p+p* collisions at midrapidity at $(s)^{1/2}$=7 and 8 TeV at the LHC have been made in Ref. [42]. The improved advanced hydrodynamic estimates [42] of initial energy density in minimum bias *p+p* collisions at midrapidity at the LHC proved to be $\varepsilon$(7 TeV)=0.645 GeV/fm³ and $\varepsilon$(8 TeV)=0.641 GeV/fm³, respectively, which are below of the critical energy density ($\approx$ 1 GeV/fm³) from lattice QCD calculations. These estimates have been obtained using the average values of the charged particle multiplicity density at midrapidity in minimum bias *p+p* collisions at $(s)^{1/2}$=7 and 8 TeV at the LHC ALICE and CMS experiments. It has been concluded [42] that there is large enough initial energy density to produce a non-hadronic (QGP-like) perfect fluid in high multiplicity *p+p* collisions at the LHC energies. The similar advanced estimates, using the relativistic hydrodynamic solutions in Ref. [43], have yielded the average initial energy density $\varepsilon$(13 TeV) $\approx$ 0.69 GeV/fm³ using the average value of the charged-particle multiplicity density <*dN*$_{ch}$/*d$\eta$*> $\approx$ 6.5 in minimum bias *p+p* collisions at $(s)^{1/2}$=13 TeV at midrapidity at the LHC. Assuming the proportionality of the initial energy density to the charged-particle multiplicity density and using results of Ref. [43], the critical value of <*dN*$_{ch}$/*d$\eta$*> at midrapidity in *p+p* collisions at $(s)^{1/2}$=13 TeV, required to attain the critical QCD energy density ($\approx$ 1 GeV/fm³), has been calculated in Ref. [21] as follows:

$$< \frac{dN_{ch}}{d\eta} > (p+p \text{ at } 13 \text{ TeV}) = \frac{1 \, GeV/fm^3}{0.69 \, GeV/fm^3} \cdot 6.5 \approx 9.4 \qquad (8).$$

Similarly, using results of Ref. [42], we can estimate the critical value of $<dN_{ch}/d\eta>$ at midrapidity in $p+p$ collisions at $(s)^{1/2}=7$ TeV, required to achieve the critical QCD energy density ($\approx 1$ GeV/fm$^3$):

$$<\frac{dN_{ch}}{d\eta}>(p+p \text{ at 7 TeV}) = \frac{1\ GeV/fm^3}{0.645\ GeV/fm^3} \cdot 5.9 \approx 9.1 \qquad (9).$$

Our estimate $<dN_{ch}/d\eta> \approx 6.1\pm0.3$ for probable deconfinement phase transition proved to be below of the critical value of $<dN_{ch}/d\eta>$ for attaining the critical energy density ($\approx 1$ GeV/fm$^3$), estimated above (Eq. (9)) for $p+p$ collisions at $(s)^{1/2}=13$ TeV at midrapidity. It is an interesting result that $<dN_{ch}/d\eta> \approx 6.1\pm0.3$ for probable deconfinement phase transition estimated in present work for $p+p$ collisions at $(s)^{1/2}=7$ TeV proved to be noticeably smaller than corresponding $<dN_{ch}/d\eta> \approx 7.1\pm0.2$ estimated in Ref. [21] for $p+p$ collisions at $(s)^{1/2}=13$ TeV, which is also seen from Figs. 4(*a*) and 4(*b*). Using the results of expressions (8) and (9), in present analysis we can estimate the respective critical energy densities for probable deconfinement phase transition in $p+p$ collisions at $(s)^{1/2}=7$ and 13 TeV corresponding to $<dN_{ch}/d\eta> \approx 6.1\pm0.3$ and $<dN_{ch}/d\eta> \approx 7.1\pm0.2$, respectively:

$$\varepsilon_{tr}(p+p \text{ at 7 TeV}) = \frac{6.1}{9.1} \cdot 1\frac{GeV}{fm^3} \approx 0.67 \pm 0.03\ \frac{GeV}{fm^3}, \qquad (10)$$

and

$$\varepsilon_{tr}(p+p \text{ at 13 TeV}) = \frac{7.1}{9.4} \cdot 1\frac{GeV}{fm^3} \approx 0.76 \pm 0.02\ \frac{GeV}{fm^3} \qquad (11).$$

As seen from above calculations, the estimated critical energy densities for probable deconfinement phase transitions in $p+p$ collisions at $(s)^{1/2}=7$ and 13 TeV proved to be significantly smaller than the critical energy density ($\approx 1$ GeV/fm$^3$) from lattice QCD calculations. From comparison of non-extensivity parameter $q$ versus $<dN_{ch}/d\eta>$ dependencies for two collision energies in Fig. 3(*a*), it has been deduced that the systems produced in $p+p$ collisions at $(s)^{1/2}=7$ TeV are characterized by noticeably larger degree of equilibrium and thermalization as compared to that in $p+p$ collisions at $(s)^{1/2}=13$ TeV, which could be due to the faster and more violent $p+p$ collisions at $(s)^{1/2}=13$ TeV than those at $(s)^{1/2}=7$ TeV. The deduced larger degree of equilibrium and thermalization at $(s)^{1/2}=7$ TeV, compared to that at $(s)^{1/2}=13$ TeV, can probably explain our finding that the probable deconfinement phase transition in $p+p$ collisions at $(s)^{1/2}=7$ TeV takes place at lower estimated critical energy density ($0.67\pm0.03$ GeV/fm$^3$) than that ($0.76\pm0.02$ GeV/fm$^3$) in $p+p$ collisions at $(s)^{1/2}=13$ TeV.

The further comprehensive analysis of multiplicity dependencies of particle production in $p+p$ collisions with the detailed investigation of various QGP signatures is necessary to determine more precisely the critical value of <dN<sub>ch</sub>/dη> and corresponding critical energy density for expected deconfinement phase transition in high-energy p+p collisions at the LHC.

## 5. Summary and Conclusions

We have analyzed the midrapidity $p_t$ spectra of the charged pions and kaons, protons and antiprotons at ten different classes of the average charged-particle multiplicity density *<dN<sub>ch</sub>/dη>* in inelastic p+p collisions at $(s)^{1/2}$=7 TeV at midrapidity at the LHC, measured by ALICE Collaboration. We have studied the evolution of collective characteristics of collision system with changing *<dN<sub>ch</sub>/dη>* by means of combined (simultaneous) minimum $\chi^2$ model fits of the $p_t$ distributions of the identified charged particles in each class of charged-particle multiplicity density, using the thermodynamically consistent Tsallis distribution function and Hagedorn function with the embedded transverse (radial) flow over measured long $p_t$ ranges of particles. The combined minimum $\chi^2$ fits with thermodynamically consistent Tsallis function as well as Hagedorn function with the embedded transverse flow describe quite well the $p_t$ spectra of identified charged particles in ten different classes of charged-particle multiplicity in p+p collisions at $(s)^{1/2}$=7 TeV. The results of the present analysis for inelastic p+p collisions at $(s)^{1/2}$=7 TeV have been compared systematically with the corresponding results of recent work [21] for inelastic p+p collisions at $(s)^{1/2}$=13 TeV.

The values of non-extensivity parameter $q$ for the charged pions and kaons, protons and antiprotons in p+p collisions at $(s)^{1/2}$=7 TeV proved to be noticeably smaller as compared to those in p+p collisions at $(s)^{1/2}$=13 TeV in the whole analyzed *<dN<sub>ch</sub>/dη>* range. This indicates that the systems produced in p+p collisions at $(s)^{1/2}$=7 TeV have the noticeably larger degree of equilibrium and thermalization compared to those in p+p collisions at $(s)^{1/2}$=13 TeV.

The extracted effective temperatures $T$ of thermodynamically consistent Tsallis function have demonstrated consistent rise with increasing the multiplicity of charged particles in p+p collisions at $(s)^{1/2}$=7 TeV in agreement with the similar result [21] obtained in p+p collisions at $(s)^{1/2}$=13 TeV. The corresponding $T$ versus *<dN<sub>ch</sub>/dη>* dependence in p+p collisions at $(s)^{1/2}$=7 TeV has been described quite well by the simple power function $T = A \cdot < \frac{dN_{ch}}{d\eta} >^\alpha$ with the same value ≈1/3 of exponent parameter as that obtained [21] in p+p collisions at $(s)^{1/2}$=13 TeV. From comparison of relation $T \sim \varepsilon^{1/3}$ deduced in present analysis and in Ref. [21] with the relation $T \sim \varepsilon_\pi^{1/4}$ for the model of an ideal gas of massless pions, it is found that the energy density dependencies of the effective temperatures of the systems, created in p+p collisions at $(s)^{1/2}$=7 TeV and 13 TeV, and that of an ideal pion gas are close to each other due to closeness of the respective exponent parameters. The difference between exponent parameter (1/3) obtained in present analysis and that (1/4) for an ideal gas of massless pions could probably be explained by nonzero viscosity to entropy ratio $((\eta/s) > 0)$ and some nonzero coupling existing between the constituents of the dense medium produced in high-energy $p+p$ collisions at the LHC.

It is obtained that the transverse (radial) flow emerges at $\langle dN_{ch}/d\eta \rangle \approx 6$ and then increases, becoming significant at higher multiplicity events and attaining the maximum value $\langle \beta_t \rangle = 0.29 \pm 0.02$ at the largest average charged-particle multiplicity density $\langle dN_{ch}/d\eta \rangle = 21.3 \pm 0.6$ in $p+p$ collisions at $(s)^{1/2}=7$ TeV. These results are consistent with the similar emergence and development of transverse radial flow in higher charged-particle multiplicity events in $p+p$ collisions at $(s)^{1/2}=13$ TeV at the LHC demonstrated in Ref. [21].

We have estimated from analysis of $T_0$ and $\langle \beta_t \rangle$ versus $\langle dN_{ch}/d\eta \rangle$ dependencies, obtained using Hagedorn function with the embedded transverse flow, that the probable deconfinement phase transition in inelastic $p+p$ collisions at $(s)^{1/2}=7$ TeV occurs at $\langle dN_{ch}/d\eta \rangle \approx 6.1 \pm 0.3$, which is noticeably smaller of the corresponding estimate ($\langle dN_{ch}/d\eta \rangle \approx 7.1 \pm 0.2$), obtained recently in $p+p$ collisions at $(s)^{1/2}=13$ TeV in Ref. [21]. We have also estimated the corresponding critical energy densities for probable deconfinement phase transitions in $p+p$ collisions at $(s)^{1/2}=7$ and 13 TeV at the LHC, which proved to be $0.67 \pm 0.03$ GeV/fm$^3$ and $0.76 \pm 0.02$ GeV/fm$^3$, respectively, being significantly lower of the critical energy density ($\approx 1$ GeV/fm$^3$) from lattice QCD calculations. The deduced noticeably larger degree of equilibrium and thermalization at $(s)^{1/2}=7$ TeV than that at $(s)^{1/2}=13$ TeV could probably explain our finding that the probable deconfinement phase transition in $p+p$ collisions at $(s)^{1/2}=7$ TeV occurs at the lower estimated critical energy density ($0.67 \pm 0.03$ GeV/fm$^3$) as compared to that ($0.76 \pm 0.02$ GeV/fm$^3$) in $p+p$ collisions at $(s)^{1/2}=13$ TeV.

**Acknowledgements**

We are thankful to ALICE publication committee for providing the tables of the published data on the measured $p_t$ distributions of particles in $p+p$ collisions at $(s)^{1/2}=7$ TeV at the LHC.

Kh.K.O. and coauthors from PTI (Tashkent) acknowledge the support provided by the Ministry of Innovative Development of Uzbekistan within the fundamental project on analysis of open data on $p+p$ and heavy-ion collisions at the RHIC and LHC. F.H.L. acknowledges the support by the Shanxi Provincial Natural Science Foundation under Grant No. 201901D111043 and Grant Nos. 11575103, 12047571, and 11947418 of the National Natural Science Foundation of China.